\documentclass[twocolumn,prl,amsmath,amssymb,superscriptaddress]{revtex4}
\usepackage{amsmath,bbm,mathrsfs}
\usepackage{graphicx}
\usepackage{amsfonts}
\usepackage{amssymb}

\def\be{\begin{equation}}
\def\ee{\end{equation}}
\def\tr{\mbox{tr}}
\def\one{\openone}
\def\compl{\mathbb{C}}

\def\textbf#1{{\bf #1}}
\def\be{\begin{equation}}
\def\ee{\end{equation}}
\def\ben{\begin{eqnarray}}
\def\een{\end{eqnarray}}
\def\eea{\end{array}}
\def\bea{\begin{array}}

\newcommand{\bei}{\begin{itemize}}
\newcommand{\eei}{\end{itemize}}
\newcommand{\ket}[1]{|#1\rangle}
\newcommand{\bra}[1]{\langle#1|}

\newcommand{\proj}[1]{\ket{#1}\!\bra{#1}}

\def\hcal{{\cal H}}

\begin{document}

\newcommand{\eg}{{\it{e.g.,~}}}
\newcommand{\ie}{{\it{i.e.,~}}}
\newcommand{\etal}{{\it{et al.}}}

\title{Unified Framework for Correlations in Terms of Local Quantum Observables}

\author{A. Ac\'in}
\affiliation{ICFO--Institut de Ci\`encies Fot\`oniques, E--08860
Castelldefels, Barcelona, Spain} \affiliation{ICREA--Instituci\'o
Catalana de Recerca i Estudis Avan\c{c}ats, Lluis Companys 23,
08010 Barcelona, Spain}
\author{R. Augusiak}
\affiliation{ICFO--Institut de Ci\`encies Fot\`oniques, E--08860 Castelldefels, Barcelona, Spain}
\author{D. Cavalcanti}
\affiliation{ICFO--Institut de Ci\`encies Fot\`oniques, E--08860
Castelldefels, Barcelona, Spain}
\author{C. Hadley}
\affiliation{ICFO--Institut de Ci\`encies Fot\`oniques, E--08860 Castelldefels, Barcelona, Spain}
\author{J.~K.~Korbicz}
\affiliation{Institute of Theoretical Physics and Astrophysics,
University of Gda\'nsk, 80-952 Gda\'nsk, Poland}
\affiliation{National Quantum Information Center of Gda\'nsk,
81-824 Sopot, Poland}
\author{M. Lewenstein}
\affiliation{ICFO--Institut de Ci\`encies Fot\`oniques, E--08860 Castelldefels, Barcelona, Spain} \affiliation{ICREA--Instituci\'o
Catalana de Recerca i Estudis Avan\c{c}ats, Lluis Companys 23, 08010 Barcelona, Spain}
\author{Ll. Masanes}
\affiliation{ICFO--Institut de Ci\`encies Fot\`oniques, E--08860 Castelldefels, Barcelona, Spain}
\author{M. Piani}
\affiliation{Institute for Quantum Computing and Department of
Physics and Astronomy, University of Waterloo, N2L 3G1 Waterloo,
Ontario, Canada}

\begin{abstract}
We provide a unified framework for nonsignalling quantum and
classical multipartite correlations, allowing all to be written as
the trace of some local (quantum) measurements multiplied by an
operator. The properties of this operator define the corresponding
set of correlations.We then show that if the theory is such that
all local quantum measurements are possible, one obtains the
correlations corresponding to the extension of Gleason's Theorem
to multipartite systems. Such correlations coincide with the
quantum ones for one and two parties, but we prove the existence
of a gap for three or more parties.
\end{abstract}

\maketitle

{\it Introduction.}---Physical principles impose limits on the
correlations observed by distant parties. It is known, for
instance, that the principle of no-signalling---that is, the fact
that the correlations cannot lead to any sort of instantaneous
communication---implies no-cloning~\cite{nocl} and
no-broadcasting~\cite{nobroad} theorems, and the possibility of
secure key distribution~\cite{qkd}. Moving to the quantum domain,
the main goal of quantum information theory is precisely to
understand how quantum properties may be used for information
processing. It is then important to understand how the quantum
formalism constrains the correlations amongst distant parties. For
instance, an asymptotically convergent hierarchy of necessary
conditions for some correlations to have a quantum realization has
been introduced in Ref.~\cite{NPA} (see also Ref.~\cite{sdp}). All
these conditions provide nontrivial bounds to the set of quantum
correlations.

The standard scenario when studying correlations consists of $N$
distant parties, $A_1,\ldots,A_N$, who can perform $m$ possible
measurements, each with $r$ possible outcomes, on their local
systems. Denote by $x_1,\ldots,x_N$ the measurement applied by the
parties and by $a_1,\ldots,a_N$ the obtained outcomes. The
observed correlations are described by the joint probability
distribution $P(a_1,\ldots,a_N|x_1,\ldots,x_N)$, giving the
probability that the parties obtain the outcomes $a_1,\ldots,a_N$
when performing the measurements $x_1,\ldots,x_N$. In full
generality, $P(a_1,\ldots,a_N|x_1,\ldots,x_N)$ is an arbitrary
vector of $m^N\times r^N$ positive entries satisfying the
normalization conditions $\sum_{a_1,\ldots,a_N}
P(a_1,\ldots,a_N|x_1,\ldots,x_N)=1$ for all $x_1,\ldots,x_N$.
These objects however become nontrivial if one wants them to be
compatible with a physical principle.

Indeed, imposing that the observed correlations should not
contradict the no-signalling principle, requires that the marginal
probability distribution observed by a group of parties, say the
first $k$ parties, be independent of the measurements performed by
the remaining $N-k$ parties. Non-signalling correlations, then,
are such that
\begin{eqnarray}\label{nscorr}
    \sum_{a_{k+1},\ldots,a_N}
    P(a_1,\ldots,a_N|x_1,\ldots,x_N)=&&\nonumber\\
    P(a_1,\ldots,a_k|x_1,\ldots,x_k),&&
\end{eqnarray}
for any splitting of the $N$ parties into two groups.

Assume now that the correlations have a quantum origin, \ie they
can be established by performing local measurements on a
multipartite quantum state. Precisely
\begin{equation}\label{qcorr}
    P_\mathrm{Q}
        =
        \{
        \tr(\rho\, M_{a_1}^{x_1} \otimes\cdots\otimes
            M_{a_N}^{x_N})
        \},
\end{equation}
where $\rho$ is a positive operator of unit trace acting on a
composite Hilbert space
$\hcal_{A_1}\otimes\ldots\otimes\hcal_{A_N}$, while
$M_{a_i}^{x_i}$ are positive operators in each local space $i$
defining the $m$ local measurements, \ie $\sum_{a_i}
M_{a_i}^{x_i}=\one_{A_i},\forall x_i$. It is well known that the
set of nonsignalling correlations is strictly larger than the
quantum set~\cite{PR}.

Finally, there is the set of classical correlations, which may be
established by sharing classically correlated data, denoted
$\lambda$. These correlations may be written in the form
\begin{equation}\label{ccorr}
    P_\mathrm{C}=
    \big\{\sum_\lambda P(\lambda)
    D_{A_1}(a_1|\lambda,x_1)\cdots D_{A_N}(a_N|\lambda,x_N)\big\} ,
\end{equation}
where $\{D_{A_i}\}$ are deterministic functions specifying the
local results of party $i$ as a function of the corresponding
measurement and the shared classical data $\lambda$. The
celebrated Bell theorem implies that the set of quantum
correlations is strictly larger than the classical
one~\cite{Bell}.

In this work, we provide a unified framework for all these sets of
correlations in terms of local quantum observables. Indeed, we
show that each of these sets of correlations can be written in the
form
\begin{equation}\label{unform}
    P_\mathrm{O} = \{\tr(O\, M_{a_1}^{x_1} \otimes\cdots\otimes M_{a_N}^{x_N})\} ,
\end{equation}
where the operators $M_{a_i}^{x_i}$ correspond to local quantum
measurements. By modifying the properties of $O$, it is possible
to generate the different sets of correlations. Requiring that
proper probabilities be derived from all possible local quantum
measurements imposes that the operator $O$ be positive on all
product states.  Namely, it must be an entanglement witness, $O=W$
\cite{footnote}. We then show that while the corresponding set of
correlations, denoted $P_W$, is equivalent to the quantum set for
one and two parties, a gap appears for $N>2$. An implication of
this result is that the extension of Gleason's Theorem to local
quantum measurements does not lead to quantum correlations for an
arbitrary number of parties.

{\it Non-signalling correlations.}---Let us start by showing how
to write any nonsignalling probability distribution in the form of
Eq. \eqref{unform} with a {\it particular} fixed set of
measurements. This is the content of the following theorem.

\textbf{Theorem 1:}~{\it An $N$-partite probability distribution
$P(a_1,\ldots ,a_N| x_1,\ldots ,x_N)$ is nonsignalling if, and
only if, there exist local quantum measurements $M_{a_i}^{x_i}$
and an Hermitian operator $O$ of unit trace such that
Eq.~\eqref{unform} holds.}

Note that the operator $O$ need not give positive probabilities
for other  measurements.

{\it Proof:} The ``if'' part is trivial, since the marginal
distributions $\sum_{a_1,\ldots,a_k}\tr(O M_{a_1}^{x_1}
\otimes\cdots\otimes M_{a_N}^{x_N})$ are clearly independent of
$x_1,\ldots,x_k$. For the ``only if'' part, we show how to obtain
$O$ and $M_{a_i}^{x_i}$ for each non-signalling distribution
$P(a_1,\ldots ,a_N| x_1,\ldots ,x_N)$.

We start by constructing the local measurements, which may be
taken, without loss of generality, to be the same for each of the
$N$ parties. First we take $m(r-1)$ vectors
$\ket{\alpha_a^x}\in\compl^d$, with $a=1,\ldots,r-1$ and
$x=1,\ldots,m$ such that the matrices $\proj{\alpha_a^x}$ and the
identity $\one$ are linearly independent, as elements of the space
of $d\times d$ Hermitian matrices. This is always possible by
taking a large enough value of the dimension $d$, \eg
$d=\max(r,m)$. Now we choose a set of positive numbers $z_a^x >0$
such that, for each value of $x$, the matrix defined as
\begin{equation}\label{lastr}
        M_{r}^x = I- \sum_{a=1}^{r -1} z_a^x\, \proj{\alpha_a^x}
\end{equation}
is positive semidefinite. This can always be achieved by choosing
sufficiently small $z_a^x$. For each value of $x$, the matrices
$M_a^x = z_a^x\proj{\alpha_a^x}$ (for $a<r$) and $M_r^x$
(\ref{lastr}) constitute a local measurement.

The set of $m(r-1)+1$ linearly independent matrices
$\{\one,M_a^x:a=1,\ldots, r-1; x=1,\ldots, m\} = \{M_1,
M_2,\ldots\}$ has a dual set $\{\tilde M_1, \tilde M_2,\ldots\}$,
such that $\tr(M_i\tilde M_j)=\delta_{ij}$. Then, the explicit
construction of the operator $O$ for the case $N=2$ is
\begin{eqnarray}
  O &=& \sum_{a_1 ,a_2 =1}^{r -1} \sum_{x_1 ,x_2 =1}^{m}
  P(a_1 ,a_2|x_1,x_2)\, \tilde{M}_{a_1}^{x_1} \otimes \tilde{M}_{a_2}^{x_2} \hspace{8mm}
  \\ \nonumber && +\
  \sum_{a_1 =1}^{r -1} \sum_{x_1 =1}^{m}
  P(a_1 |x_1)\, \tilde{M}_{a_1}^{x_1} \otimes \tilde{\one}
  \\ \nonumber && +\
  \sum_{a_2 =1}^{r -1} \sum_{x_2 =1}^{m}
  P(a_2 |x_2)\, \tilde{\one}\otimes \tilde{M}_{a_2}^{x_2}
  \,+\, \tilde{\one}\otimes \tilde{\one}\ .
\end{eqnarray}
The marginal probabilities $P(a_1|x_1)$ and $P(a_2|x_2)$ are well
defined because $P(a_1 ,a_2|x_1,x_2)$ is nonsignalling. Note that,
since the dual matrices $\tilde{M}^x_a$ are, in general, not
positively defined, neither is $O$. After some simple algebra, one
can see that this operator and the previous local measurements
reproduce the initial probability distribution according to Eq.
\eqref{unform}. It also follows directly from the construction,
that $O$ is Hermitian and $\tr(O)=1$.

The generalization to higher $N$ is based on the fact that
nonsignalling distributions are characterized by the numbers
$P(a_1,\ldots, a_N| x_1,\ldots, x_N)$ for $a_i <r$, together with
all the $(N-1)$-party marginals (e.g. $P(a_2,\ldots, a_N|
x_2,\ldots, x_N)$). These marginals, being themselves
non-signaling distributions, are also characterized by the entries
with $a_i <r$, plus all the $(N-2)$-party marginals. Recursively,
one arrives at the single-party marginals, which by normalization,
are characterized by the entries with $a_i <r$ too. $\square$

As an illustration of the formalism, we give the explicit form of
the operator $O$ and local measurements reproducing the
Popescu--Rohrlich correlations (or ``PR-box''~\cite{PR}).  This
represents the best known example of nonsignalling correlations
not attainable by quantum means, in which the algebraic maximum of
the Clauser--Horne--Shimony--Holt inequality~\cite{CHSH} is
achieved. This distribution is defined to be $P_{\rm PR}(a,b|x,y)
= 1/2$ if $xy = a+b\mod 2$, and $0$ otherwise, where $a,b,x,y$ are
now bits. In this case, the required operator, which is surely not
an entanglement witness, reads $O = \alpha^+\Phi^+ +
\alpha^-\Phi^-$, where $\Phi^\pm$ are the projectors onto the Bell
states $|\Phi^\pm\rangle=(1/\sqrt{2})(\ket{00}\pm\ket{11})$, and
$\alpha^\pm = (1\pm\sqrt{2})/2$; and the local observables are
$\{\sigma_x,\sigma_y\}$ for Alice, and $\{(\sigma_x -
\sigma_y)/\sqrt 2,(\sigma_x + \sigma_y)/\sqrt 2\}$ for Bob.

This theorem induces a hierarchical structure for the different
sets of correlations. By constraining the form of $O$, it is
possible to generate the sets of quantum and classical
correlations. Indeed, one can encapsulate all the previous sets of
correlations in the following statement.

The distribution  $P(a_1,\ldots,a_N|x_1,\ldots,x_N)$ is
\begin{itemize}
    \item {\it Nonsignalling} if, and only if, it may be written in the form of Eq. \eqref{unform};
    \item {\it Quantum} whenever the operator $O$ is positive;
    \item {\it Local} if, and only if, $O$ corresponds to a separable quantum state~\cite{AGM}.
\end{itemize}

{\it Gleason's Theorem for local quantum observables.}---Consider
now a theory such that all possible local {\it quantum}
measurements are allowed. In this case, the operator $O$ is
required to be positive on all product states, implying that it
must be an {\it entanglement witness} $W$ with $\tr(W)=1$. Thus,
the corresponding set of correlations reads
\begin{equation}\label{wcorr}
    P_\mathrm{W}=\{\tr(W M_{a_1}^{x_1}\otimes\cdots\otimes M_{a_N}^{x_N})\}.
\end{equation}
Since $W$ needs not be positive, the set of
correlations~\eqref{wcorr} could be larger than the quantum set.

Interestingly, these correlations have already appeared in several
works studying the extension of Gleason's Theorem to local
observables in $N$ independent Hilbert spaces. In what follows, we
name the set of correlations defined by Eq.~\eqref{wcorr} as {\it
Gleason correlations}. Recall that Gleason's Theorem is a
celebrated result in quantum mechanics proving that any map from
generalized measurements to probability distributions can be
written as the trace rule with the appropriate quantum
state~\cite{gleason}. More precisely: Let ${\cal P(H)}$ be the set
of operators $M$ acting on ${\cal H}$ such that $0 \leq M \leq
\one$. For any map $v: {\cal P(H)} \rightarrow [0,1]$ such that
$\sum_i v(M_i) = 1$ when $\sum_i M_i=\one$, 
there is a positive operator $\rho$ such that $v(M) = \tr(\rho
M)$. A simple proof of this theorem can be found in
Ref.~\cite{Busch}. This theorem has been generalized to the case
of local observables acting on bipartite~\cite{localgleason} and
general multipartite~\cite{wallach} systems. In the same fashion,
as the original theorem, the goal is now to characterize those
maps from $N$ measurements---one for each party---to joint
nonsignalling probability distributions. It has been shown in
these works that for each of these maps, there is a witness
operator $W$ such that $v(M_1,\ldots,M_N) = \tr(W M_1
\otimes\ldots\otimes M_N)$.

{\bf Theorem 2:}~{\it There exist Gleason correlations
$P_{\mathrm{W}'}=\mathrm{tr} (W'\, \Pi_{a_1}^{x_1}\otimes
\Pi_{a_2}^{x_2}\otimes \Pi_{a_3}^{x_3})$ (\ie obtained by applying
local projective measurements $\Pi_{a_i}^{x_i}$ on a normalized
entanglement witness $W'$) that do not have a quantum realization,
\ie such that there exist no quantum state $\rho$ and local
measurements, $M_{a_i}^{x_i}$, in an arbitrary tripartite Hilbert
space satisfying $P_{\mathrm{W}'}=\mathrm{tr}(\rho
M_{a_1}^{x_1}\otimes M_{a_2}^{x_2}\otimes M_{a_3}^{x_3})$.}

{\it Proof.} To show that the set of Gleason correlations is
strictly larger than the quantum set, we construct a witness and
local measurements which lead to a violation of a Bell inequality
higher than the quantum one.

The Bell inequality we consider has been introduced in
Ref.~\cite{almeida} for the tripartite scenario in which the
parties apply two measurements each with two possible outcomes. We
label the choice of measurements and the obtained results by bits.
The inequality reads
\begin{equation}\label{mafin}
    \beta=p(000|000)+p(110|011)+p(011|101)+p(101|110)\leq 1 .
\end{equation}
One can indeed prove that the values achievable through classical
and quantum means are at most unity; that is, the inequality is
not violated by quantum theory~\cite{almeida}.

Moving to the definition of the operator $W'$, we consider the
witness which detects the three-qubit bound entangled state of
Ref.~\cite{upb} based on unextendible product bases (UPB). Recall
that an unextendible product basis in a composite Hilbert space of
total dimension $d$ is defined by a set of $n<d$ orthogonal
product states which cannot be completed into a full product
basis, as there is no other product state orthogonal to them. In
Ref.~\cite{upb} an example of such a set of product states for
three qubits was constructed. It consists of the following four
states:
\begin{equation}\label{upb3q}
\ket{000},\quad\ket{1e^\bot
e},\quad\ket{e1e^\bot},\quad\ket{e^\bot e1}
\end{equation}
where $\{\ket e,\ket{e^\bot}\}$ is an arbitrary basis different
from the computational one. We denote by $\Pi_\mathrm{UPB}$ the
projector onto the subspace spanned by these states. One knows
that the state $\rho_\mathrm{UPB}=(\one-\Pi_\mathrm{UPB})/4$ is
bound entangled. A normalized witness $W'$ detecting this state is
given by
\begin{equation}\label{witness}
    W'=\frac{1}{4-8\epsilon}(\Pi_\mathrm{UPB}-\epsilon\one),
\end{equation}
where $\epsilon=\min_{\ket{\alpha \beta \gamma}}\bra{\alpha \beta
\gamma}\Pi_\mathrm{UPB}\ket{\alpha \beta \gamma}$. One immediately
confirms that here $0<\epsilon<1/2$. Clearly, $W'$ is positive on
all product states and detects $\rho_\mathrm{UPB}$, since
$\tr(W'\rho_\mathrm{UPB})=-\epsilon/4(1-2\epsilon)$ which is
negative for any $\epsilon<1/2$.

Now, one can see that the witness $W'$ when measured in the local
bases in the definition of the UPB~\eqref{upb3q} leads to
correlations such that
\begin{equation}
    \beta=\frac{1-\epsilon}{1-2\epsilon} ,
\end{equation}
which is larger than unity for $0<\epsilon<1/2$. $\square$

This theorem, then, proves that the set of Gleason correlations is
strictly larger than the quantum set for $N>2$. The equivalence of
these two sets in the bipartite scenario $N=2$ has recently been
shown in~\cite{caltech}. For the sake of completeness, we present
here a slightly simpler proof of this result.

The Choi--Jamio\l{}kowski (CJ) isomorphism implies that any
witness $W$ can be written as
$(\one_{A_1}\otimes\Upsilon_{A_2})(\Phi)$, where $\Upsilon$ is a
positive map and $\Phi$ is the projector onto the maximally
entangled state. Using the same techniques as in
Ref.~\cite{HHHContr}, one can prove that any witness can also be
written as $(\one_{A_1}\otimes\Lambda_{A_2})(\Psi)$, where
$\Lambda$ is now positive and trace-preserving and $\Psi$ is a
projector onto a pure bipartite state. Denoting by $\Lambda^{*}$
the dual~\cite{notedual} of $\Lambda$, we have 
\begin{eqnarray}\label{bipgleason}
    \tr(W M_{a_1}^{x_1}\otimes M_{a_2}^{x_2})&=&
    \tr[(\one\otimes\Lambda)(\Psi) M_{a_1}^{x_1}\otimes M_{a_2}^{x_2}]\nonumber\\
    &=&\tr[\Psi M_{a_1}^{x_1}\otimes\Lambda^* (M_{a_2}^{x_2})]\nonumber\\
    &=&
    \tr(\Psi M_{a_1}^{x_1}\otimes\tilde M_{a_2}^{x_2}) ,
\end{eqnarray}
%
where $\tilde M_{a_2}^{x_2}=\Lambda^*(M_{a_2}^{x_2})$ defines a
valid quantum measurement because the dual of a positive
trace-preserving map is positive and unital, \ie,
$\Lambda^*(\one)=\one$.

{\it Discussion.}---There is an ongoing effort to understand the
gap between quantum and nonsignalling correlations. As stated
above, there exist correlations which, despite being compatible
with the no-signalling principle, cannot be attained by local
measurements on a quantum state. The natural question is then to
study why these supra-quantum correlations do not seem to be
observed in nature. Of course, a trivial answer to this question
is that there exist no positive operator and projective
measurements in a Hilbert space reproducing these supra-quantum
correlations via the Born trace rule. However, one would wish for
a set of ``natural'' principles with which to exclude these
supra-quantum correlations. These principles would provide a
better, or ideally the exact, characterization of quantum
correlations.

Most of the principles proposed so far to rule out supra-quantum
correlations have an information theoretic motivation. The idea is
that the existence of these correlations would imply an important
change in the way information is processed and transmitted. It has
been shown, for instance, that communication complexity would
become trivial if the PR-box, or some noisy version of it, were
available~\cite{commcompl}, that some of these supra-quantum
correlations violate a new information principle called {\it
information causality}~\cite{singapore}, or that they would lead
to the violation of macroscopic locality~\cite{macroscopic}.
Unfortunately, none of these principles has been proven to be able
to single out the set of quantum correlations~\cite{axioms}.

In this work, we introduce a unified mathematical formalism for
nonsignalling and quantum correlations in terms of local quantum
observables. We expect this formalism to be useful when tackling
all such questions. It may be easier using our construction to
study how new constraints may be added to the nonsignalling
principle in order to derive the quantum correlations. The methods
developed here may also be useful to study the degree of
non-locality of quantum states, witnesses, and $O$-operators.

We have, then, considered Gleason correlations, defined by
nonsignalling theories in which all possible local quantum
measurements are possible. We have shown the presence of a gap
between this set and the quantum set of correlations for $N>2$
parties. Thus, while the hypothesis in Gleason's Theorem for local
observables completely characterizes the set of bipartite quantum
correlations~\cite{caltech}, the result does not extend to the
multipartite scenario. Clearly, the proof of equivalence in the
bipartite case exploits the existence of the CJ isomorphism.
Actually, it is easy to see that the equivalence holds for those
$N$-party entanglement witnesses that can be written
\begin{equation}
    W=\sum_k p_{k}\left[\Lambda^k_{A_1}\otimes\cdots\otimes
    \Lambda^k_{A_N}\right](\rho_k) ,
\end{equation}
where $\rho_k$ are $N$-party quantum states, $p_{k}$ some
probability distribution, and $\Lambda^k_{A_i}$ are positive,
trace-preserving maps and the number of terms in the sum is
arbitrary. Our results imply that this decomposition is not
possible for all $N$-party entanglement witnesses. It would be
interesting to better understand why the theorem fails in the
multipartite scenario and identify additional requirements able to
close the gap.

We thank J. Barrett, S. Boixo, E. Cavalcanti, D. Chru\'sci\'nski,
S. Pironio, G. Sarbicki for discussions. J. K. K. and M. P.
acknowledge ICFO for kind hospitality. This work was supported by
the Spanish MEC/MINCIN projects TOQATA (FIS2008-00784), QTIT
(FIS2007-60182) and QOIT (Consolider Ingenio 2010), EU Integrated
Project QAP and SCALA and STREP NAMEQUAM, ERC Grants QUAGATUA and
PERCENT, Caixa Manresa, Generalitat de Catalunya, ICFO--OCE
collaborative programs, Alexander von Humboldt Foundation,
QuantumWorks, and OCE.

\bibliographystyle{plain}

\begin{thebibliography}{99}
\bibitem{footnote} Recall that an
entanglement witness is a non-positive Hermitian operator such
that $\bra{\alpha,\beta\ldots}W\ket{\alpha,\beta,\ldots}\geq 0$
for all product states $\ket{\alpha,\beta,\ldots}$. We omit here
the non-positivity requirement so that the witnesses form a convex
set that contains the convex set of quantum states, which in turn
contains the convex set of separable states.
\bibitem{notedual}
The dual map $\Lambda^*$ of $\Lambda$ is the map such that
$\tr[A\Lambda(B)]=\tr[\Lambda^*(A)B]$ for all $A$ and $B$.
\end{thebibliography}

\end{document}